\documentclass[aps,pra,reprint]{revtex4-2}

\usepackage[utf8]{inputenc}
\usepackage{amsthm}
\usepackage{amsmath}
\usepackage{color}
\usepackage{hyperref}
\usepackage{graphicx,xcolor}
\usepackage{amssymb}
\usepackage{bbold}
\usepackage{gensymb}
\usepackage{tikz}

\newcommand{\ii}{\mathrm{i}} 
\newcommand{\eul}{\mathrm{e}} 
\newcommand{\diff}{\mathrm{d}} 
\newcommand{\id}{\mathbb{1}} 

\newcommand{\ket}[1]{|#1\rangle} 
\newcommand{\bra}[1]{\langle#1|} 
\newcommand{\braket}[1]{\langle#1\rangle} 
\newcommand{\ketbra}[2]{|#1\rangle\!\langle #2|} 
\newcommand{\tr}{\mathrm{tr}} 

\def\kgate at (#1,#2,#3){
\draw [rounded corners=2pt, fill={rgb,255:red,208; green,240; blue,208} , line width=0.5pt, font=\small ] (#1+0.3,#2+4) rectangle  node {#3} (#1+3.7,#2+6);
\node[font=\small] (il) at (#1+1,#2+7.5) {}; 
\draw (il) -- (#1+1, #2+6);
\node[font=\small] (i2) at (#1+3,#2+7.5) {}; 
\draw (i2) -- (#1+3, #2+6);
\node[font=\small] (il) at (#1+1,#2+2.5) {}; 
\draw (il) -- (#1+1, #2+4);
\node[font=\small] (i2) at (#1+3,#2+2.5) {}; 
\draw (i2) -- (#1+3, #2+4);
;}
\def\kgatez at (#1,#2,#3){
\draw [rounded corners=2pt, fill={rgb,255:red,208; green,240; blue,208} , line width=0.5pt, font=\small ] (#1+0.3,#2+4) rectangle  node {#3} (#1+3.7,#2+6);
\node[font=\small] (il) at (#1+1,#2+7.5) {}; 
\draw (il) -- (#1+1, #2+6);
\node[font=\small] (i2) at (#1+3,#2+7.5) {}; 
\draw (i2) -- (#1+3, #2+6);
\node[font=\small] (il) at (#1+2,#2+2.5) {}; 
\draw (il) -- (#1+2, #2+4);
;}
\def\btens at (#1,#2){
\draw [rounded corners=2pt, fill={rgb,255:red,255; green,250; blue,235} , line width=0.5pt, font=\small ] (#1-0.5,#2-0.5) rectangle  node {$x$} (#1+0.5,#2+0.5);
;}

\begin{document}

\title{Tensor network influence functionals for open quantum systems with general Gaussian bosonic baths}

\author{Valentin Link}
\affiliation{Institut für Physik und Astronomie, Technische Universität Berlin, D-10623, Berlin, Germany}

\begin{abstract}
Dynamics of open quantum systems with structured reservoirs can often be simulated efficiently with tensor network influence functionals. The standard variants of the time-evolving matrix product operator (TEMPO) method are applicable when the systems is coupled to Gaussian bosonic baths via hermitian coupling operators that mutually commute. In this work we introduce a generalization to cases where the system is coupled to a single reservoir through multiple non-commuting operators, representing the most general form of linear system-bath coupling. We construct a Gaussian influence functional that properly handles Trotter errors arising from a finite evolution time step, thus ensuring convergence for long evolution times. Based on this result, the uniform TEMPO scheme can be employed to obtain a matrix product operator form of the influence functional, enabling efficient simulations of the real-time dynamics of the open system. As a demonstration, we simulate the time evolution of driven two-level emitters coupled to a bosonic lattice at different lattice sites. 
\end{abstract}\maketitle

\section{Introduction} 

When a quantum system is strongly coupled to a reservoir, traditional concepts of open quantum system theory such as master equations and dynamical maps often no longer provide a suitable framework to characterize the local system dynamics \cite{devega}. Instead, strong memory effects inherent in the problem favor a description in terms of quantum multi-time processes, in particular process tensors and influence functionals \cite{pollock2018,ortegataberner2024,keelingProcessTensorApproaches2025}. This approach transparently reveals the structure of temporal correlations in the dynamics, and can be made practically useful through concepts from quantum many-body theory, tensor networks, and quantum information \cite{banulsMatrixProductStates2009a,leroseInfluenceMatrixApproach2021, Carignano2024}. Several approaches for numerical simulation of the open system dynamics directly employ this framework by constructing numerically feasible matrix product operator (MPO) representations of influence functionals \cite{MuellerHermes2012, jorgensenExploitingCausalTensor2019, CygorekACE}. The success of these methods relies on the structure of temporal correlations in the local dynamics of certain many body systems, which sometimes behave favorably over spatial correlations in the wave function of the full system-and-bath model \cite{friasperezLightConeTensor2022}. This holds in particular for non-interacting reservoirs, where rigorous bounds for the simulation complexity of the local dynamics have been established \cite{vilkoviskiyBoundApproximatingNonMarkovian2024, Thoenniss2025}. Given an influence functional represented as a compressed temporal MPO, real-time dynamics can be computed very efficiently. This approach has been applied to a variety of physical problems, including the computation of equilibrium spectra \cite{nguyenCorrelationFunctionsTensor2024a,sonner2025Semigroup, dewitProcessTensorApproaches2025,salamonMarkovianApproachNPhoton2026, garbelliniUniformProcessTensor2026}, non-equilibrium dynamics \cite{mickiewiczExactFloquetDynamics2025, kahlertSimulatingLandauZener2024}, transport \cite{fuxTensorNetworkSimulation2023}, quantum control \cite{fuxEfficientExplorationHamiltonian2021, ortegataberner2024}, quantum thermodynamics \cite{shubrookNumericallyExactOpen2025} and quantum information \cite{dowlingCapturingLongRangeMemory2024, backerVerifyingQuantumMemory2025a}.

Tensor network algorithms such as the time-evolving matrix product operator (TEMPO) method \cite{strathearnEfficientNonMarkovianQuantum2018, jorgensenExploitingCausalTensor2019} automatically generate a MPO representation of the influence functional in a controlled way. They can be applied to structured reservoirs at low temperatures \cite{nguyenCorrelationFunctionsTensor2024a, sonner2025Semigroup}, such as encountered, for instance, in impurity problems in solid-state systems \cite{kotliarElectronicStructureCalculations2006} or complex chemical environments \cite{nuominEfficientSimulationOpen2024}. This makes the approach promising for systems that are coupled to a highly coherent bath via multiple channels. However, even for Gaussian baths with linear coupling, obtaining exact expressions for influence functionals on a discretized time grid that can be used in a tensor network compression algorithm is a non-trivial task. For instance, in the case of fermionic baths one has to solve the non-interacting (single particle) evolution in order to determine the exact memory kernel of the influence functional \cite{ngRealtimeEvolutionAnderson2023,sonner2025Semigroup}, whereas using the text-book continuum path integral result yields large errors due to time discretization (Trotter errors) \cite{nayakSteadystateDynamicalMean2025a}. 
On the other hand, continuum algorithms that avoid time-axis discretizations are at an early stage of development and are currently not competitive \cite{parkTensorNetworkInfluence2024, wassnerFormalisingOperationalContinuum2026}.

In this technical contribution we provide an elementary derivation of a time-discrete ``Trotterized'' influence functional for the case of multiple non-commuting coupling operators that may couple to the same bath modes. This most general scenario is typically not addressed in the literature. Previously, commuting hermitian coupling operators \cite{strathearnEfficientNonMarkovianQuantum2018} or non-commuting hermitian couplings to individual reservoirs \cite{palmQuasiadiabaticPathIntegral2018,gribbenExactDynamicsNonadditive2022a} have been considered, with the exception of Ref.~\cite{richter2022Enhanced} which uses a perturbative expansion of the influence functional. This alternative representation leads to a non-Gaussian form that induces additional technical overhead in simulations. In contrast, the approach presented here retains the Gaussian structure of the bath response, resembling the usual Feynman-Vernon expression with minimal but non-trivial corrections that assure consistent convergence with respect to the time-step size. Therefore, this result can be used directly for dynamical simulations with the available TEMPO algorithms. We focus on the case of stationary baths with finite memory time, such that the efficient uniform TEMPO scheme \cite{linkOpenQuantumSystem2024a} can be applied. 

This paper is structured as follows. We first introduce the general Gaussian bosonic bath model in Sec.\ref{sec:bath}. Then the Trotterized influence functional is defined in Sec.~\ref{sec:IF} and its analytical expression is provided. In Sec.~\ref{sec:conver} we show simple numerical tests for spin-boson models with Jaynes-Cummings type coupling, demonstrating consistent convergence of dynamics simulations. In Sec.~\ref{sec:exampl} we apply the method to a challenging example problem, two driven emitters coupled to a bosonic lattice, showing the current capabilities and limitations of the method. Lastly we summarize our conclusions in Sec.~\ref{sec:concl}. 

\section{Gaussian bath models}\label{sec:bath}

We consider the most general model of a stationary Gaussian bosonic reservoir coupled linearly to an arbitrary system
\begin{equation}
    H=H_\mathrm{sys}\otimes \id_\mathrm{env}+ \sum_{l=1}^L S^l \otimes B^l +\id_\mathrm{sys}\otimes H_\mathrm{env}.
\end{equation}
Here, $L$ is the number of distinct hermitian coupling operators $S^l$ of the system that need not mutually commute $[S^l, S^m]\neq 0$. The bath operators are given as
\begin{equation}
    B^l=\sum_k(h_{k}^lb_k+h_{k}^{l*}b_k^\dagger),\qquad H_\mathrm{env}=\sum_k \omega_{k} b_{k}^\dagger b_{k}.
\end{equation}
The annihilation operators $b_k$ describe bosonic bath modes with a stationary Gaussian initial state $\rho_\mathrm{env}(0)$, where stationary means explicitly that $[H_\mathrm{env},\rho_\mathrm{env}(0)]=0$. The stationarity condition is not strictly necessary for the derivation of the influence functional, but it resembles the standard scenario including thermal reservoirs. We define the bath correlation function as
\begin{equation}\label{eq:bcf_general}
    \alpha^{lm}(t)=\mathrm{tr}_\mathrm{env} \eul^{\ii H_\mathrm{env}t}B^l \eul^{-\ii H_\mathrm{env}t}B^m\rho_\mathrm{env}(0).
\end{equation}
This two-time correlation function completely characterizes the influence of the Gaussian reservoir onto the system dynamics \cite{devega, mascherpaOptimizedAuxiliaryOscillators2020}. We moreover assume without loss of generality $\mathrm{tr}_\mathrm{env} B^m\rho_\mathrm{env}(0)=0$.
In the common scenario one considers a single coupling term $L=1$. Then the bath correlation function for a thermal environment at temperature $k_\mathrm{B}T=1/\beta$ is a single function (not a matrix) and reads explicitly
\begin{equation}\label{eq:bcf_single}
\begin{split}
        \alpha(t)&=\sum_k |h_k|^2 \left((1+n_\mathrm{B}(\omega_k))\eul^{-\ii\omega_k t}+n_\mathrm{B}(\omega_k)\eul^{\ii\omega_k t}\right)\\
        &=\int\limits_0^\infty \diff\omega J(\omega)\left((1+n_\mathrm{B}(\omega))\eul^{-\ii\omega t}+n_\mathrm{B}(\omega)\eul^{\ii\omega t}\right)
\end{split}
\end{equation}
with the thermal Bose occupation number $n_\mathrm{B}(\omega)=(\eul^{\beta\omega}-1)^{-1}$ and the spectral density
\begin{equation}\label{eq:spectral_density}
    J(\omega)=\sum_k |h_k|^2\delta(\omega-\omega_k),
\end{equation}
which is assumed to be a continuous function. For this problem the exact time-discretized influence functional is well-known and closely resembles the continuum expression \cite{makriTensorPropagatorIterative1995, palmQuasiadiabaticPathIntegral2018}. However, already the simple Janynes-Cummings-type coupling
\begin{equation}
H_\mathrm{int}=\sum_k h_k (C^\dagger b_k+ C b_k^\dagger) +\sum_k \omega_k b_k^\dagger b_k
\end{equation}
no longer falls into this class when the system operator $C$ is non-hermitian. This coupling structure often arises in quantum optics from performing the rotating wave approximation. Note that this case can still be treated in a straightforward manner using hierarchical equations of motion (HEOM) \cite{tanimuraNumericallyExactApproach2020,HierarchicalEOMjl2023}. However, in HEOM it is difficult to describe more general cross-correlations in the bath correlation function, such as those that arise when multiple subsystems couple to a spatially extended reservoir at different positions. We proceed with defining a time-discretized influence functional for the general case \eqref{eq:bcf_general}.

\section{Time-discrete influence functional}\label{sec:IF}
For the following derivations we introduce a Fourier representation of the bath correlation function
\begin{equation}
    \alpha^{lm}(t)=\sum_\lambda g_\lambda^l g_\lambda^{m*}\eul^{-\ii\omega_\lambda t},
\end{equation}
where $\omega_\lambda$ are an extended set of frequencies taking positive and negative values (usually a continuum, here written in a discretized form for convenience). This decomposition always exists because the bath correlation function is a positive kernel. We can then consider a substitute Hamiltonian with
\begin{equation}\label{eq:hamilt}
\begin{split}
    &H_\mathrm{int}(t)=\sum_{l=1}^L S^l\otimes B^l(t)\\ & B^l(t)=\sum_\lambda (g_\lambda^{l*} b_\lambda^\dagger \eul^{\ii \omega_\lambda t}+ g_\lambda^{l} b_\lambda \eul^{-\ii \omega_\lambda t}).
\end{split}
\end{equation}
This new environment has the same Gaussian bath response as the original environment, but with a zero temperature initial condition $\rho_\mathrm{aux}(0)=\ketbra{0}{0}$. Note that one has to allow for modes with negative frequency $\omega_\lambda<0$ that can describe absorption of energy from the bath. It is easy to check that
\begin{equation}
    \braket{0|B^l(t)B^m(s)|0}=\alpha^{lm}(t-s).
\end{equation}
This decomposition is used merely as a convenience for the derivation of the influence functional and need not be explicitly determined, as the influence functional depends fundamentally only on $\alpha$ itself. We first define the unitary evolution operator generated by $H_\mathrm{int}$ via
\begin{equation}
    \partial_tU(t,s)=-\ii H_\mathrm{int}(t)U(t),\qquad U(s,s)=\id.
\end{equation}
We then employ a symmetric Trotter decomposition for every coupling term in the Hamiltonian
\begin{equation}\label{eq:Utrotter}
\begin{split}
            &{U}(t+2\delta t,t)\approx \\&{U}^1(t+2\delta t,t+\delta t)\cdots {U}^L(t+2\delta t,t+\delta t)\\&\cdot{U}^L(t+\delta t,t)\cdots {U}^1(t+\delta t,t),
\end{split}\end{equation}
 assuming a small time step $\delta t$.
The individual unitaries $U^l$ are generated, respectively, by the $L$ coupling terms in $H_\mathrm{int}$. The symmetric Trotterization is of central importance here. It ensures that the time-step error will be suppressed sufficiently such that evolution to long times can be brought to convergence (see Fig.~\ref{fig:trotter_conv}). Note that every $U^l$ is diagonal in the eigenbasis of the respective system coupling operator $S^l$. Defining the different eigenbases for every coupling operator via $S^l\ket{i,l}=S^l_i\ket{i,l}$ we have the following decomposition
\begin{equation}
U ^l(t,s)=\sum_{i=1}^d \ketbra{i,l}{i,l}\otimes U^{l}_i (t,s),
\end{equation}
where the conditioned operators $U_i^l(t,s)$ act on the environment Hilbert space
\begin{equation}
    U^{l}_i (t,s)=\braket{i,l|U^l (t,s)|i,l}.
\end{equation}
In the case of degeneracies in the spectrum of the operator $S^l$ the sum can be restricted to the distinct eigenvalues when the projectors $P^l_i=\ketbra{i,l}{i,l}$ are replaced with projectors on the $r_l\leq d$ distinct eigenspaces, i.e.
\begin{equation}
\begin{split}\label{eq:U_decomp}
    &U ^l(t,s)=\sum_{i=1}^{r_l} P^l_{i}\otimes U^{l}_i (t,s) 
\end{split}.
\end{equation}
This degeneracy filtering reduces the index dimensions of the influence functional, which is important for numerical performance. 
The conditioned propagators $U^{l}_i (t,s)$ are Gaussian unitaries on the bath Hilbert space and they obey the evolution equation
\begin{equation}
    \partial_tU^l_i(t,s)=-\ii S_i^l\sum_\lambda (g_\lambda^l \eul^{-\ii\omega_\lambda t}b_\lambda+g_\lambda^{l*} \eul^{\ii\omega_\lambda t}b_\lambda^\dagger ) U^{l}_i(t,s),
\end{equation}
where $S^l_i$ is the $i$'th eigenvalue of $S^l$. An influence functional can then be defined in the following way. 
\begin{widetext}
Considering $H_\mathrm{sys}=0$ for now, the reduced system state after $N$ (assumed even) time steps is given as
\begin{equation}\label{eq:rho_sys}
    \rho_\mathrm{sys}(N\delta t)= \tr_\mathrm{env}\!\left[U^{1}(N\delta t,N\delta t-\delta t)\cdots U^{L}(\delta t,0)\cdots U^{1 }(\delta t,0)\ket{0}\rho_\mathrm{sys}(0)\bra{0}U^{1\dagger}(\delta t,0)\cdots U^{L\dagger }(N\delta t,N\delta t- \delta t)\right].
\end{equation}
In the language of Keldysh path-integrals, the unitaries appearing to the left of $\rho_0$ define the ``forward'' contour whereas the conjugated unitaries to the right give rise to the ``backward'' contour \cite{KamenevBook}. Inserting the decomposition \eqref{eq:U_decomp} and collecting all terms involving the bath one can define the influence functional
\begin{equation}\label{eq:full_if}
\mathcal{F}^{i_1^1 \ldots i_1^Li_2^L\ldots i_N^1}_{j_1^1 \ldots j_1^Lj_2^L\ldots j_N^1}=\bra{{0}}U^{1\,\dagger}_{j_1^1}(\delta t,0)\cdots U^{L\,\dagger}_{j_1^L}(\delta t,0)\cdots U^{1\,\dagger}_{j_{N}^L}(N\delta t,N\delta t- \delta t)U^{1}_{i_N^L}(N\delta t,N\delta t-\delta t)\cdots U^{1 }_{i_1^1}(\delta t,0)\ket{{0}}.
\end{equation}
Here we use system basis indices $i^l_n$ for the ``forward'' unitaries and $j^l_n$ for the ``backward'' (conjugated) unitaries, where $l$ and $n$ denote the coupling term and the time step, respectively.
This expression can be evaluated in an elementary way exploiting that the unitaries $U^l_i$ are simple displacement operators that map coherent states to coherent states. The full derivation provided in Appendix \ref{app:deriv}. Up to Trotter corrections in the time-local term, the influence functional resembles the Feynman-Vernon result
\begin{equation}
\begin{split}\label{eq:infl_gauss}
        &\mathcal{F}^{i_1^1 \ldots i_1^Li_2^L\ldots i_N^1}_{j_1^1 \ldots j_1^Lj_2^L\ldots j_N^1}=
        \exp\left(-\sum_{n=1}^N\sum_{m=1}^{n-1}\sum_{l,o=1}^L(S^{l}_{i_n^l}-S^{l}_{j_n^l}) (\eta^{lo}_{n-m}S^{o}_{i_m^{o}}-(\eta^{lo}_{n-m})^*S^{o}_{j_m^{o}})\right)\left(\mathcal{F}_0\right)^{i_1^1 \ldots i_1^Li_2^L\ldots i_N^1}_{j_1^1 \ldots j_1^Lj_2^L\ldots j_N^1},
\end{split}
\end{equation}
\end{widetext}
with the discretized bath correlation function
\begin{equation}
    \eta_k^{lo}=\int_{(k-1)\delta t}^{k\delta t}\diff t \int_{0}^{\delta t}\diff s  \alpha^{lo}(t-s)
\end{equation}
and $\mathcal{F}_0$ containing only ``Markovian'' (time-local) contributions. Note that this same expression holds also in the case of commuting couplings $[S^l,S^m]=0$ irrespective of the Trotter decomposition. However, the symmetric decomposition induces nontrivial corrections in the time-local term, given as 
\begin{widetext}
\begin{equation}
\begin{split}\label{eq:infl_gauss0}
        &\ln\left(\left(\mathcal{F}_0\right)^{i_1^1 \ldots i_1^Li_2^L\ldots i_N^1}_{j_1^1 \ldots j_1^Lj_2^L\ldots j_N^1}\right)=\\&
        \sum_{n=1}^N\sum_{l,o=1}^L\Big(-R^n_{l-o}S^{l}_{i_n^l}\eta^{lo}_{0}S^{o}_{i_n^{o}}-S^{l}_{i_n^l}(\tilde\eta^{lo}_{0})^*S^{o}_{j_n^{o}}-S^{l}_{j_n^l}\tilde\eta^{lo}_{0}S^{o}_{i_n^{o}}+R^n_{l-o}S^{l}_{j_n^l}(\eta^{lo}_{0})^*S^{o}_{j_n^{o}}+\delta_{lo}S^{l}_{i_n^l}\tilde\eta^{ll}_{0}S^{l}_{i_n^{o}}+\delta_{lo}S^{l}_{j_n^l}(\tilde\eta^{ll}_{0})^*S^{o}_{j_n^{o}}\Big)
\end{split}
\end{equation}    
\end{widetext}
with 
\begin{equation}
     \tilde\eta_0^{lo}=\int_{0}^{\delta t}\diff t \int_{0}^t\diff s\alpha^{lo}(t-s).
\end{equation}
We moreover defined the matrix $R^n_k$ that discriminates even and odd time steps, in accordance with the symmetric Trotter decomposition
\begin{equation}
    R^n_k=\begin{cases}
        \Theta_k &n \text{ odd}\\
        \Theta_{-k} &n \text{ even}
    \end{cases}\qquad \Theta_k=\begin{cases}
        0 &k\leq 0 \\
        1 &k >0
    \end{cases}.
\end{equation}
The corrections in Eq.~\eqref{eq:infl_gauss0} are a central result of this work. 
Note that this representation of the influence functional is directly amenable to all standard variants of the quasi-adiabatic path integral (QUAPI) and TEMPO algorithms. In Sec.~\ref{sec:unitempo} we recapitulate the uniform TEMPO scheme \cite{linkOpenQuantumSystem2024a}, which shows superior numerical scaling over finite-time algorithms. UniTEMPO generates a semi-group embedding of the dynamics such that evolution to long times is directly possible without additional extrapolative expansions \cite{ cerrilloNonMarkovianDynamicalMaps2014a, makriSmallMatrixDisentanglement2020}. Moreover, the algorithm is computationally cheap compared to QUAPI \cite{makriTensorPropagatorIterative1995} or previous versions of TEMPO \cite{jorgensenExploitingCausalTensor2019} and can be used for very long bath memory times. The core algorithm provides a factorization of the influence functional as a matrix product operator in the form
\begin{equation}\label{eq:IFMPO}
    \mathcal{F}^{i_1^1 \ldots i_1^Li_2^L\ldots i_N^1}_{j_1^1 \ldots j_1^Lj_2^L\ldots j_N^1}=\vec{v}_l^T (f_\mathrm{e})^{i_N^1 \ldots i_N^L}_{j_N^1 \ldots j_N^L}\cdots (f_\mathrm{o})^{i_1^1 \ldots i_1^L}_{j_1^1 \ldots j_1^L}\vec{v}_r
\end{equation}
where $f_\mathrm{e/o}$ are alternating, time-local tensors for even and odd time steps and are independent of the number of time steps $N$. The corresponding tensor elements are square matrices
\begin{equation}
    (f_\mathrm{e/o})^{i_n^1 \ldots i_n^L}_{j_n^1 \ldots j_n^L}\in \mathbb{C}^{\chi \times \chi},\qquad \vec{v}_{l/r}\in \mathbb{C}^\chi,
\end{equation}  
and $\chi$ is the MPO bond dimension. This dimension determines the accurarcy of the representation, where large $\chi$ corresponds to high accuracy but large numerical effort. The boundary vectors $\vec{v}_{l/r}$ live in the bond space and effectively realize the bath initial state and trace (see Eq.~\eqref{eq:qevolution}).
Given such a factorization of the influence functional in terms of a matrix product one can define an effective propagator for the non-Markovian dynamics. For this we need to go back to the original expression \eqref{eq:rho_sys}. The influence functional \eqref{eq:full_if} is related to the system evolution through the local projectors $P^l_i$ from Eq.~\eqref{eq:U_decomp}. 

From now on we switch to a graphical notation in order to avoid lengthy formulas. In this common notation tensors are written as boxes and indices are outgoing ``legs'' (lines). If the legs of two tensors are connected a summation is implied, see Ref.~\cite{keelingProcessTensorApproaches2025} for an introduction in an open systems context. We write the local system basis projectors $P^l_i = \ketbra{i,l}{i,l}$ graphically as
\begin{equation}
    \begin{tikzpicture}[inner sep=1mm, x=.23cm,y=.23cm,every node/.style={scale=1}]
    \node[] (x) at (-8, 2) {$\bra{\phi}P^l_i\ket{\psi} =$};
    \node[] (A) at (1,4) {$i$};
    \draw (1,1) -- (A);
    \node[] (x) at (-2.5, 1) {$\bra{\phi} $};
    \node[] (z) at (4.5, 1) {$\ket{\psi} $};
    \draw (1,1) -- (x);
    \draw (1,1) -- (z);
    \draw [rounded corners=5pt, fill={rgb,255:red,170; green,227; blue,255} , line width=0.5pt, font=\small ] (0,0) rectangle  node {$P^l$} (2,2);
    
    \end{tikzpicture}.
\end{equation}
These projectors appear on both the forward and backward contour, so it is convenient to use the double notation
\begin{equation}
        \begin{tikzpicture}[inner sep=1mm, x=.23cm,y=.23cm,every node/.style={scale=1}]
    \begin{scope}[shift={(1,0.5)}]
    \node[] (A) at (1,4) {$j$};
    \draw (1,1) -- (A);
    \node[] (x) at (-2.5, 1) {};
    \node[] (z) at (4.5, 1) {};
    \draw (1,1) -- (x);
    \draw (1,1) -- (z);
    \draw [rounded corners=5pt, fill={rgb,255:red,152; green,221; blue,240} , line width=0.5pt, font=\small ] (0,0) rectangle  node {} (2,2);
    \end{scope}
    
    \node[] (x) at (-7, 2) {$P^l_i\otimes P^l_j \,=$};
    \node[] (A) at (1,4) {$i$};
    \draw (1,1) -- (A);
    \node[] (x) at (-2.5, 1) {};
    \node[] (z) at (4.5, 1) {};
    \draw (1,1) -- (x);
    \draw (1,1) -- (z);
    \draw [rounded corners=5pt, fill={rgb,255:red,170; green,227; blue,255} , line width=0.5pt, font=\small ] (0,0) rectangle  node {$P^l$} (2,2);
    
    \end{tikzpicture}.
\end{equation}
We further write the elementary uniTEMPO tensors $f_\mathrm{e/o}$ as
\begin{equation}\label{eq:ftens1}
    \begin{tikzpicture}[inner sep=1mm, x=.23cm,y=.23cm,every node/.style={scale=1}]
        \draw[line width=1] (-1.5,5) -- (11.5,5);
    \node[] at (-6.5,5) {$(f_\mathrm{o})^{i_n^1 \ldots i_n^L}_{j_n^1 \ldots j_n^L}\,=$};
    \draw [rounded corners=2pt, fill={rgb,255:red,255; green,225; blue,225} , line width=0.5pt, font=\small ] (0,4) rectangle  node {$f_\mathrm{o}$} (10,6);
    \node[font=\small] (il) at (1,8) {$j_n^L$}; 
    \draw (il) -- (1, 6);
    \node[font=\small] (i2) at (7,8) {$j_n^2$}; 
    \draw (i2) -- (7, 6);
    \node[font=\small] (i1) at (9,8) {$j_n^1$}; 
    \draw (i1) -- (9, 6);
    \node[font=\small] (il) at (1,2) {$i_n^L$}; 
    \draw (il) -- (1, 4);
    \node[font=\small] (i2) at (7,2) {$i_n^2$}; 
    \draw (i2) -- (7, 4);
    \node[font=\small] (i2) at (9,2) {$i_n^1$}; 
    \draw (i2) -- (9, 4);

    \node[font=\scriptsize] at (4,8) {$\cdots$};
    \node[font=\scriptsize] at (4,2) {$\cdots$};
    \end{tikzpicture},
\end{equation}
\begin{equation}\label{eq:ftens2}
\begin{tikzpicture}[inner sep=1mm, x=.23cm,y=.23cm,every node/.style={scale=1}]
        \draw[line width=1] (-1.5,5) -- (11.5,5);
    \node[] at (-6.5,5) {$(f_\mathrm{e})^{i_n^1 \ldots i_n^L}_{j_n^1 \ldots j_n^L}\,=$};
    \draw [rounded corners=2pt, fill={rgb,255:red,255; green,225; blue,225} , line width=0.5pt, font=\small ] (0,4) rectangle  node {$f_\mathrm{e}$} (10,6);
    \node[font=\small] (il) at (1,8) {$j_n^1$}; 
    \draw (il) -- (1, 6);
    \node[font=\small] (i2) at (3,8) {$j_n^{2}$}; 
    \draw (i2) -- (3, 6);
    \node[font=\small] (i1) at (9,8) {$j_n^L$}; 
    \draw (i1) -- (9, 6);
    \node[font=\small] (il) at (1,2) {$i_n^{1}$}; 
    \draw (il) -- (1, 4);
    \node[font=\small] (i2) at (3,2) {$i_n^{2}$}; 
    \draw (i2) -- (3, 4);
    \node[font=\small] (i2) at (9,2) {$i_n^L$}; 
    \draw (i2) -- (9, 4);

    \node[font=\scriptsize] at (6,8) {$\cdots$};
    \node[font=\scriptsize] at (6,2) {$\cdots$};
\end{tikzpicture},
\end{equation}
where the open legs attached to the left and right of the tensor box correspond to the matrix index (MPO bond) of dimension $\chi$.
The projectors can be combined with the tensors $f$ in order to define a propagator for a double time step $2\delta t$ via
\begin{equation}\label{eq:q_def}
\begin{tikzpicture}[inner sep=1mm, x=.23cm,y=.23cm,every node/.style={scale=1}]
    \node[] at (-17,3) {$q=$};
    \draw[line width=1] (-3,5) -- (11.5,5);
    \begin{scope}[shift={(1,0.5)}]
    \node[] (A) at (1,5) {};
    \draw (1,1) -- (A);
    \node[] (x) at (-2, 1) {};
    \node[] (z) at (4, 1) {};
    \draw (1,1) -- (x);
    \draw (1,1) -- (z);
    \draw [rounded corners=5pt, fill={rgb,255:red,152; green,221; blue,240} , line width=0.5pt, font=\small ] (0,0) rectangle  node {} (2,2);
            \draw [] plot [smooth, tension=1.5] coordinates {(1,3) (0.65,6.25) (0,5.5)};

    \end{scope}
    \node[] (A) at (1,5) {};
    \draw (1,1) -- (A);
    \node[] (x) at (-2, 1) {};
    \node[] (z) at (4, 1) {};
    \draw (1,1) -- (x);
    \draw (1,1) -- (z);
\draw [rounded corners=5pt, fill={rgb,255:red,170; green,227; blue,255} , line width=0.5pt, font=\small ] (0,0) rectangle  node {$P^{\!L}$} (2,2);
    \begin{scope}[shift={(8.2,0)}]
        [inner sep=1mm, x=.23cm,y=.23cm,every node/.style={scale=1}]
    \begin{scope}[shift={(1,0.5)}]
    \node[] (A) at (1,5) {};
    \draw (1,1) -- (1,3);
    \node[] (x) at (-2, 1) {};
    \node[] (z) at (4, 1) {};
    \draw (1,1) -- (x);
    \draw (1,1) -- (z);
    \draw [rounded corners=5pt, fill={rgb,255:red,152; green,221; blue,240} , line width=0.5pt, font=\small ] (0,0) rectangle  node {} (2,2);
    \draw [] plot [smooth, tension=1.5] coordinates {(1,3) (0.65,6.25) (0,5.5)};
    \end{scope}
    \node[] (A) at (1,5) {};
    \draw (1,1) -- (A);
    \node[] (x) at (-2, 1) {};
    \node[] (z) at (4, 1) {};
    \draw (1,1) -- (x);
    \draw (1,1) -- (z);
\draw [rounded corners=5pt, fill={rgb,255:red,170; green,227; blue,255} , line width=0.5pt, font=\small ] (0,0) rectangle  node {$P^{1}$} (2,2);
    \end{scope}
    
    \draw [rounded corners=2pt, fill={rgb,255:red,255; green,225; blue,225} , line width=0.5pt, font=\small ] (0,4) rectangle  node {$f_\mathrm{o}$} (10,6);

    \node[font=\scriptsize] at (6.1,1.5) {$\cdots$};
    \node[font=\scriptsize] at (5.1,1) {$\cdots$};
    \node[font=\scriptsize] at (5,6.8) {$\cdots\cdots$};
    \node[font=\scriptsize] at (5,3.2) {$\cdots\cdots$};

\begin{scope}[shift={(-13,0)}]
        \draw[line width=1] (-1.5,5) -- (13,5);
    \begin{scope}[shift={(1,0.5)}]
    \node[] (A) at (1,5) {};
    \draw (1,1) -- (A);
    \node[] (x) at (-2, 1) {};
    \node[] (z) at (4, 1) {};
    \draw (1,1) -- (x);
    \draw (1,1) -- (z);
    \draw [rounded corners=5pt, fill={rgb,255:red,152; green,221; blue,240} , line width=0.5pt, font=\small ] (0,0) rectangle  node {} (2,2);
            \draw [] plot [smooth, tension=1.5] coordinates {(1,3) (0.65,6.25) (0,5.5)};

    \end{scope}
    \node[] (A) at (1,5) {};
    \draw (1,1) -- (A);
    \node[] (x) at (-2, 1) {};
    \node[] (z) at (4, 1) {};
    \draw (1,1) -- (x);
    \draw (1,1) -- (z);
\draw [rounded corners=5pt, fill={rgb,255:red,170; green,227; blue,255} , line width=0.5pt, font=\small ] (0,0) rectangle  node {$P^{1}$} (2,2);
    \begin{scope}[shift={(8.2,0)}]
        [inner sep=1mm, x=.23cm,y=.23cm,every node/.style={scale=1}]
    \begin{scope}[shift={(1,0.5)}]
    \node[] (A) at (1,5) {};
    \draw (1,1) -- (1,3);
    \node[] (x) at (-2, 1) {};
    \node[] (z) at (4, 1) {};
    \draw (1,1) -- (x);
    \draw (1,1) -- (z);
    \draw [rounded corners=5pt, fill={rgb,255:red,152; green,221; blue,240} , line width=0.5pt, font=\small ] (0,0) rectangle  node {} (2,2);
    \draw [] plot [smooth, tension=1.5] coordinates {(1,3) (0.65,6.25) (0,5.5)};
    \end{scope}
    \node[] (A) at (1,5) {};
    \draw (1,1) -- (A);
    \node[] (x) at (-2, 1) {};
    \node[] (z) at (4, 1) {};
    \draw (1,1) -- (x);
    \draw (1,1) -- (z);
\draw [rounded corners=5pt, fill={rgb,255:red,170; green,227; blue,255} , line width=0.5pt, font=\small ] (0,0) rectangle  node {$P^{\!L}$} (2,2);
    \end{scope}
    
    \draw [rounded corners=2pt, fill={rgb,255:red,255; green,225; blue,225} , line width=0.5pt, font=\small ] (0,4) rectangle  node {$f_\mathrm{e}$} (10,6);

    \node[font=\scriptsize] at (6.1,1.5) {$\cdots$};
    \node[font=\scriptsize] at (5.1,1) {$\cdots$};
    \node[font=\scriptsize] at (5,6.8) {$\cdots\cdots$};
    \node[font=\scriptsize] at (5,3.2) {$\cdots\cdots$};

\end{scope}
\end{tikzpicture}
\end{equation}
capturing the full propagation due to the system-bath interaction. We can include a non-vanishing system Hamiltonian in the standard way by using another Trotter decomposition. We define the local system evolution channel
\begin{equation}
        \begin{tikzpicture}[inner sep=1mm, x=.23cm,y=.23cm,every node/.style={scale=1}]
    \begin{scope}[shift={(1,0.5)}]
    \node[] (x) at (-2.5, 1) {};
    \node[] (z) at (4.5, 1) {};
    \draw (1,1) -- (x);
    \draw (1,1) -- (z);
    \draw [rounded corners=5pt, fill={rgb,255:red,152; green,221; blue,240} , line width=0.5pt, font=\small ] (0,0) rectangle  node {} (2,2);
    \end{scope}
    
    \node[] (x) at (-10, 1.5) {$\eul^{-\ii H_\mathrm{sys}\delta t}\otimes \eul^{\ii H_\mathrm{sys}\delta t} = 
$};
    \node[] (x) at (-2.5, 1) {};
    \node[] (z) at (4.5, 1) {};
    \draw (1,1) -- (x);
    \draw (1,1) -- (z);
    \draw [rounded corners=5pt, fill={rgb,255:red,170; green,227; blue,255} , line width=0.5pt, font=\small ] (0,0) rectangle  node {$u$} (2,2);
    
    \end{tikzpicture}
\end{equation}
assuming a time-independent Hamiltonian for simplicity. Combining this local unitary with the tensor $q$ from Eq.~\eqref{eq:q_def} the system density matrix can be evolved in time via iterative applications of interaction and system propagators. The state after $2N$ time steps can be explicitly computed as
\begin{equation}\label{eq:qevolution}
    \begin{tikzpicture}[inner sep=1mm, x=.23cm,y=.23cm,every node/.style={scale=1}]
    \node[] at (1.7,3) {$\begin{pmatrix}
        & \hspace{7em} \\
        & \\
        & \\
        & \\
        &
    \end{pmatrix}^{\!\!\! N} $};
    \draw[line width=1] (-7,5) -- (11,5);
    \draw (-2,1) -- (4.5,1);
    \draw (-2,1.5) -- (4.5,1.5);
    \draw [rounded corners=2pt, fill={rgb,255:red,255; green,225; blue,225} , line width=0.5pt, font=\small ] (0,0) rectangle  node {$q$} (2.5,6);
    \begin{scope}[shift={(4,0)}]
    \begin{scope}[shift={(1,0.5)}]
    \node[] (x) at (-2, 1) {};
    \node[] (z) at (7, 1) {};
    \draw (1,1) -- (x);
    \draw (1,1) -- (z);
    \draw [rounded corners=5pt, fill={rgb,255:red,152; green,221; blue,240} , line width=0.5pt, font=\small ] (0,0) rectangle  node {} (2,2);
    \end{scope}
    \node[] (x) at (-2, 1) {};
    \node[] (z) at (8, 1) {};
    \draw (1,1) -- (x);
    \draw (1,1) -- (z);
    \draw [rounded corners=5pt, fill={rgb,255:red,170; green,227; blue,255} , line width=0.5pt, font=\small ] (0,0) rectangle  node {$u$} (2,2);
    \end{scope}

    \begin{scope}[shift={(-4.5,0)}]
    \begin{scope}[shift={(1,0.5)}]
    \node[] (x) at (-4, 1) {};
    \node[] (z) at (4, 1) {};
    \draw (1,1) -- (x);
    \draw (1,1) -- (z);
    \draw [rounded corners=5pt, fill={rgb,255:red,152; green,221; blue,240} , line width=0.5pt, font=\small ] (0,0) rectangle  node {} (2,2);
    \end{scope}
    \node[] (x) at (-3.5, 1) {};
    \node[] (z) at (4, 1) {};
    \draw (1,1) -- (x);
    \draw (1,1) -- (z);
    \draw [rounded corners=5pt, fill={rgb,255:red,170; green,227; blue,255} , line width=0.5pt, font=\small ] (0,0) rectangle  node {$u$} (2,2);
    \end{scope}
        \draw [rounded corners=8pt, fill={rgb,255:red,200; green,255; blue,255} , line width=0.5pt, font=\small ] (11,0) rectangle  node {$\rho_0$} (13.5,2.5);
        \draw [rounded corners=8pt, fill={rgb,255:red,255; green,225; blue,225} , line width=0.5pt, font=\small ] (11,4) rectangle  node {$\vec{v}_r$} (13.5,6.5);
        \draw [rounded corners=8pt, fill={rgb,255:red,255; green,225; blue,225} , line width=0.5pt, font=\small ] (-9.5,4) rectangle  node {$\vec{v}_l$} (-7,6.5);

\end{tikzpicture}.
\end{equation}
The vectors $\vec{v}_{l/r}$ from Eq.~\eqref{eq:IFMPO} can be interpreted as direct replacements of the initial environment state and the environment trace within the compressed MPO bond space. 
This formalism is not restricted to density matrix evolution. In fact, all local multi-time correlations can be extracted in a similar way using the MPO representation of the influence functional via the effective propagator defined in Eq.~\eqref{eq:qevolution} \cite{salamonMarkovianApproachNPhoton2026, garbelliniUniformProcessTensor2026}. In fact, this propagator defines a so-called process tensor capturing the full multi-time physics of the system \cite{backerVerifyingQuantumMemory2025a}.

\section{Uniform TEMPO}\label{sec:unitempo}

In this section the details of the uniform TEMPO scheme (see also Refs.~\cite{linkOpenQuantumSystem2024a,Dudgeon2025Quantum, keelingProcessTensorApproaches2025,cochinEfficientConstructionTimeinvariant2026}) are repeated using the notation from the previous section. 
UniTEMPO is based on the process tensor variant of TEMPO called PT-TEMPO \cite{jorgensenExploitingCausalTensor2019,fuxEfficientExplorationHamiltonian2021}. In this approach one first represents the influence functional as a two-dimensional tensor network and then contracts the network to MPO form, using MPO compression based in singular value decomposition (SVD) to circumvent exponential growth of the MPO bond dimension. In uniTEMPO this network is mapped to an infinite network with time-translation invariance that can be directly contracted to infinite MPO using standard infinite time-evolving block decimation (iTEBD). Detailed descriptions of the algorithm can be found in Refs.~\cite{linkOpenQuantumSystem2024a,keelingProcessTensorApproaches2025}. Note that the influence functional \eqref{eq:infl_gauss} has the same structure as in the standard Feynman-Vernon ($L=1$) case, so that this algorithm can be directly applied. The only minor difference are the Trotter corrections appearing in \eqref{eq:infl_gauss0}, which will only affect the final step of the algorithm. 

Consider for all $k>0$ the tensors
\begin{equation}\label{eq:Itens}
\begin{split}
        &(I(k))^{i^1_n\ldots i_n^Lj_n^1\ldots j_n^L }_{i^1_m\ldots i_m^Lj_m^1\ldots j_m^L}=\\&\exp\!\Big(-\sum_{l,o=1}^L(S^{l}_{i_n^l}-S^{l}_{j_n^l}) (\eta^{lo}_{n-m}S^{o}_{i_m^{o}}-(\eta^{lo}_{n-m})^*S^{o}_{j_m^{o}})\Big).
\end{split}
\end{equation}
In order to avoid notational overhead we combine the index sets $\{i^1_n,\ldots ,i_n^L,j_n^1,\ldots ,j_n^L \}$ to single multi-indices $\mu_n$. The dimension of this multi-index is $r_1^2\cdots r_L^2$, where $r_l\leq d$ denotes the number of distinct eigenvalues of the $l$'th coupling operator $S^l$.
The PT-TEMPO network is build from the following elementary tensors
\begin{equation}\label{eq:b_tens}
    \begin{tikzpicture}[inner sep=1mm, x=.23cm,y=.23cm,every node/.style={scale=1}]
    \node[] at (-10,5) {$(b(k))^{\mu\beta}_{\nu\alpha}=\delta_{\alpha\mu}\delta_{\beta\nu}(I(k))^\beta_\alpha=$};
    \draw [rounded corners=2pt, fill={rgb,255:red,208; green,240; blue,208} , line width=0.5pt, font=\small ] (0.3,4) rectangle  node {$k$} (3.7,6);
    \node[font=\small] (il) at (1,8) {$\mu$}; 
    \draw (il) -- (1, 6);
    \node[font=\small] (i2) at (3,8) {$\beta$}; 
    \draw (i2) -- (3, 6);
    \node[font=\small] (il) at (1,2) {$\nu$}; 
    \draw (il) -- (1, 4);
    \node[font=\small] (i2) at (3,2) {$\alpha$}; 
    \draw (i2) -- (3, 4);
    \end{tikzpicture}.
\end{equation}
Note that these tensors always have a degeneracy on the $\beta$ and $\nu$ indices. This is due to the casual structure of the Keldysh action in \eqref{eq:Itens}, where only the differences of eigenvalues $S^{l}_{i_n^l}-S^{l}_{j_n^l}$ (Keldysh ``quantum" field components) appear on the left of the quadratic form \cite{KamenevBook}. The degeneracy can be used to reduce the corresponding index dimensions in order to make the network contraction more efficient \cite{strathearnEfficientNonMarkovianQuantum2018}.
The analogue tensors for $k=0$ have a different structure and discriminate between even and odd steps
\begin{equation}
    \begin{tikzpicture}[inner sep=1mm, x=.23cm,y=.23cm,every node/.style={scale=1}]
\node[] at (-12,5) {$(b_\mathrm{e/o}(0))^{\mu\beta}_{\nu}=\delta_{\beta\mu}\delta_{\beta\nu}(I_\mathrm{e/o}(0))^\beta=$};
    \draw [rounded corners=2pt, fill={rgb,255:red,208; green,240; blue,208} , line width=0.5pt, font=\small ] (0.3,4) rectangle  node {$\mathrm{e/o}$} (3.7,6);
    \node[font=\small] (il) at (1,8) {$\mu$}; 
    \draw (il) -- (1, 6);
    \node[font=\small] (i2) at (3,8) {$\beta$}; 
    \draw (i2) -- (3, 6);
    \node[font=\small] (il) at (2,2) {$\nu$}; 
    \draw (il) -- (2, 4);
    \end{tikzpicture},
\end{equation}
where the definition of $I_{\mathrm{e/o}}(0)$ follows analogously from Eq.~\eqref{eq:infl_gauss0}.
We further define the ``initial'' tensor
\begin{equation}
\begin{tikzpicture}[inner sep=1mm, x=.23cm,y=.23cm,every node/.style={scale=1}]
\node[] at (-2.5,3) {$x_\mu\equiv 1=$};
\btens at (2,4.5);
\node[font=\small] (il) at (2,2) {$\mu$}; 
\draw (il) -- (2, 4);
\end{tikzpicture},
\end{equation}
which corresponds to a trivial summation over the index $\mu$. It is a straightforward exercise to show that the influence functional $\mathcal{F}$ can be expressed exactly via the network (see Ref.~\cite{strathearnEfficientNonMarkovianQuantum2018, jorgensenExploitingCausalTensor2019} for explicit derivations)
\begin{equation}\label{eq:TEMPO_net}
\begin{tikzpicture}[inner sep=1mm, x=.23cm,y=.23cm,every node/.style={scale=1}]

\foreach \j in {0,4,8,12,16} {
\kgate at (\j, 0, 1);
};
\foreach \j in {2,6,10,14} {
\kgate at (\j, 3, 2);
};
\foreach \j in {4,8,12} {
\kgate at (\j, 6, 3);
};
\foreach \j in {6,10} {
\kgate at (\j, 9, 4);
};
\kgate at (8, 12, 5);
\foreach \j in {-2,6,14} {
\kgatez at (\j, -3, $\mathrm{e}$);
};
\foreach \j in {2,10,18} {
\kgatez at (\j, -3, $\mathrm{o}$);
};
\btens at (-1,4);
\btens at (1,7);
\btens at (3,10);
\btens at (5,13);
\btens at (7,16);
\btens at (9,19);
\btens at (21,4);
\btens at (19,7);
\btens at (17,10);
\btens at (15,13);
\btens at (13,16);
\btens at (11,19);
\node at (0,-1) {$\mu_6$};
\node at (4,-1) {$\mu_5$};
\node at (8,-1) {$\mu_4$};
\node at (12,-1) {$\mu_3$};
\node at (16,-1) {$\mu_2$};
\node at (20,-1) {$\mu_1$};
\end{tikzpicture}
\end{equation}
where, for simplicity, we used $N=6$ time steps. In the uniTEMPO approach one considers baths with a finite memory time so that all gates $b(k)$ for $k>N_c$ can be neglected. This happens because the exponent in Eq.~\eqref{eq:Itens} includes the discretized bath correlation function $\eta_k$, which decays to zero at large $k$. When the exponent is sufficiently small then $I(k)\approx 1$ and the gate $b(k)$ becomes a simple swap gate, acting trivially on the swap-invariant initial tensors. In practice the cutoff $N_c$ is already implicitly determined by the compression tolerance, and need not be chosen manually \cite{sonner2025Semigroup}. In uniTEMPO one considers the long-time limit $N=\infty$ and contracts the network in the bulk only using conventional iTEBD. In the bulk, the time-translational invariant network then becomes
\begin{equation}\label{eq:uniTEMPO_network}
\begin{tikzpicture}[inner sep=1mm, x=.23cm,y=.23cm,every node/.style={scale=1}]
\foreach \j in {0,4,8,12,16} {
\kgate at (\j, 0, 1);
};
\foreach \j in {-2,6,14,2,10,18} {
\kgate at (\j, 3, 2);
};
\foreach \j in {0,4,8,12,16} {
\kgate at (\j, 9, $N_c\!\!-\!\!1$);
};
\foreach \j in {-2,6,14,2,10,18} {
\kgate at (\j, 12, $N_c$);
};
\foreach \j in {-2,6,14} {
\kgatez at (\j, -3, $\mathrm{e}$);
};
\foreach \j in {2,10,18} {
\kgatez at (\j, -3, $\mathrm{o}$);
};
\foreach \j in {-1,1,3,5,7,9,11,13,15,17,19,21} {
\btens at (\j,19);
};
\node at (0,-1) {$\mu_{n+5}$};
\node at (4,-1) {$\mu_{n+4}$};
\node at (8,-1) {$\mu_{n+3}$};
\node at (12,-1) {$\mu_{n+2}$};
\node at (16,-1) {$\mu_{n+1}$};
\node at (20,-1) {$\mu_n$};

\node at (-4, 19) {$\cdots$};
\node at (-4, 17) {$\cdots$};
\node at (-2, 14) {$\cdots$};
\node at (-4, 8) {$\cdots$};
\node at (-2, 5) {$\cdots$};
\node at (-4, 2) {$\cdots$};

\node at (24, 19) {$\cdots$};
\node at (24, 17) {$\cdots$};
\node at (22, 14) {$\cdots$};
\node at (24, 8) {$\cdots$};
\node at (22, 5) {$\cdots$};
\node at (24, 2) {$\cdots$};

\node at (0, 11) {$\cdots$};
\node at (4, 11) {$\cdots$};
\node at (8, 11) {$\cdots$};
\node at (12, 11) {$\cdots$};
\node at (16, 11) {$\cdots$};
\node at (20, 11) {$\cdots$};

\end{tikzpicture}.
\end{equation}
This network can be interpreted as a many-body quantum evolution of a translation invariant product initial state, with local tensor $x$ at each site, undergoing alternating next-neighbor interactions via the $b(k)$ gates, i.e.~evolution from top to bottom in \eqref{eq:uniTEMPO_network}. Computing this ``dynamics'' in reverse memory time via infinite time-evolving block decimation (iTEBD) \cite{vidalClassicalSimulationInfiniteSize2007} directly yields an infinite MPO form
\begin{equation}\label{eq:mpo_if}
\begin{tikzpicture}[inner sep=1mm, x=.23cm,y=.23cm,every node/.style={scale=1}]
\node[font=\small] (i) at (22,-2) {$\mu_{n}$}; 
\draw (i) -- (22,1);
\node[font=\small] (i) at (18,-2) {$\mu_{n+1}$}; 
\draw (i) -- (18,1);
\node[font=\small] (i) at (14,-2) {$\mu_{n+2}$}; 
\draw (i) -- (14,1);
\node[font=\small] (i) at (10,-2) {$\mu_{n+3}$}; 
\draw (i) -- (10,1);
\node[font=\small] (i) at (6,-2) {$\mu_{n+4}$}; 
\draw (i) -- (6,1);
\node[font=\small] (i) at (2,-2) {$\mu_{n+5}$}; 
\draw (i) -- (2,1);

\draw[line width=1] (-0,1) -- (24,1);
\node at (-0.9, 0.95) {$\cdots$};
\node at (25.1, 0.95) {$\cdots$};

\foreach \j in {0,8,16} {
    \draw [rounded corners=2pt, fill={rgb,255:red,255; green,225; blue,225} , line width=0.5pt, font=\small ] (\j+0.5,0) rectangle  node {$f_\mathrm{e}$} (\j+3.5,2);
};
\foreach \j in {4,12,20} {
    \draw [rounded corners=2pt, fill={rgb,255:red,255; green,225; blue,225} , line width=0.5pt, font=\small ] (\j+0.5,0) rectangle  node {$f_\mathrm{o}$} (\j+3.5,2);
};
\end{tikzpicture}
\end{equation}
defining the tensors in Eq.~\eqref{eq:ftens1}, \eqref{eq:ftens2} upon expanding the multi-indices $\mu_n=(i_n^1\ldots i_n^L;j_n^1\ldots j_n^L)$. 
Note that a distinction between even and odd timesteps happens only at the last network layer, which fuses two virtual ``sites'' and does not increase the MPO bond dimension. Thus, the Trotter corrections do not add complexity to the contraction algorithm, in contrast to the construction used in Ref.~\cite{richter2022Enhanced}. 

This bulk contraction is sufficient for realizing dynamics with certain stationary initial conditions. Physical boundary vectors $\vec{v}_{l/r}$ can be obtained as left and right fixed points of the tensor $q$ (Eq.~\eqref{eq:q_def}) and tracing out the system degrees of freedom. In order to realize the standard product initial state of system and bath, one can use the decoupling scheme introduced in Ref.~\cite{linkOpenQuantumSystem2024a}. There one increases the dimension of all indices in the tensor network by one, introducing a ``zero'' index value such that $(I(k))^0_\mu=(I(k))^\mu_0=1$. Boundary vectors can then be obtained from computing the fixed points of $f_\mathrm{e}^0=f_\mathrm{o}^0$, i.e.~$f_\mathrm{e}^0\vec{v}_{r}=\vec{v}_{r}$ and $\vec{v}_{l}^T f_\mathrm{e}^0=\vec{v}_{l}^T$. The extra zero index value is then discarded in order to construct the propagator $q$ as before. This procedure realizes the product state initial condition implied in the finite network Eq.~\eqref{eq:TEMPO_net}, as proven in Ref.~\cite{linkOpenQuantumSystem2024a}.

\section{Convergence}\label{sec:conver}

\begin{figure}
    \centering
    \includegraphics[width=225pt]{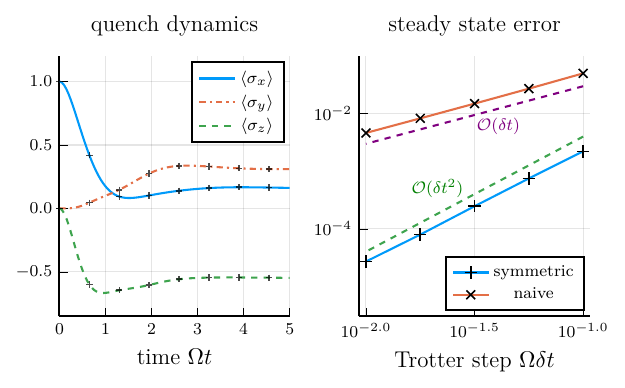}\vspace{-10pt}
    \caption{Simulation of the driven Jaynes-Cummings model with $H_\mathrm{sys}=\Omega\sigma_x/2$ and the bath \eqref{eq:bcff_lorenz} $\omega=g=\Gamma=2\Omega,\, \bar n=0.25$. Left: The dynamics computed with uniTEMPO is in perfect agreement with the reference solution (black markers) computed via the full system \eqref{eq:HJC}. Right: Convergence of the steady state with respect to the Trotter time step $\delta t$. Using the symmetric splitting \eqref{eq:Utrotter} yields quadratic error scaling, whereas the error is linear when omitting the Trotter corrections in Eq.~\eqref{eq:infl_gauss0}. Simulations were performed at bond dimension $\chi\approx 28$ ($N_c=2048$). The error is computed via the matrix norm of the difference of the steady state computed with uniTEMPO and the numerically exact reference state $||\rho-\rho_\mathrm{ref}||_2$.}
    \label{fig:trotter_conv}
\end{figure}

We now demonstrate numerically the convergence of the method with respect to the Trotter step. For this we consider a damped Jaynes-Cummings model for which the full non-Markovian dynamics can be computed directly. The model consists of a single spin coupled to a single bosonic mode
\begin{equation}\label{eq:HJC}
H=\frac{\Omega}{2}\sigma_x + g(a^\dagger \sigma_-+a\sigma_+)+\omega a^\dagger a,
\end{equation}
where the mode $a$ is additionally subjected to Markovian damping and pumping with Lindblad operators $L_1=\sqrt{\gamma(1+\bar{n})}a$, $L_2=\sqrt{\gamma \bar n}a^\dagger$. The parameters $g,\omega,\bar n$ and $\gamma$ represent the coupling strength, frequency, occupation, and decay rate of the mode. The mode is assumed to be initially in the uncoupled steady state such that its response is stationary. The bath response is also Gaussian because the Lindblad master equation contains only terms that are quadratic in the bosonic operators. In a non-Markovian description of the two-level system alone, where the mode is traced out, we can set~$S^1=\sigma_x$, $S^2=\sigma_y$ and obtain the bath correlation function
\begin{equation}\label{eq:bcff_lorenz}
\alpha(t)=\frac{1}{4}\begin{pmatrix} \alpha_+(t) & -\ii \alpha_-(t) \\ \ii \alpha_-(t) & \alpha_+(t)
\end{pmatrix},
\end{equation}
\begin{equation}
\alpha_\pm(t)={g^2} (1+\bar n)\eul^{-\ii\omega t- \gamma |t|}\pm {g^2} \bar n\eul^{+\ii\omega t-\gamma |t|}.
\end{equation}
Note that this bath model violates the fluctuation-dissipation relation implied in the form Eq.~\eqref{eq:bcf_single} and hence does not resemble a thermal reservoir. 
We can easily obtain a numerically exact reference solution by solving the full Markovian model of spin and mode. This can be used to verify the solution obtained from simulating the non-Markovian reduced dynamics via the non-commuting generalization of uniTEMPO. In Fig.~\ref{fig:trotter_conv} we display an example quench dynamics, verifying the uniTEMPO simulation. We moreover display the convergence of the steady state in order to assess accuracy at long evolution times. We confirm quadratic scaling of the steady state error with the time step $\delta t$. In contrast, naively using the influence functional from Ref.~\cite{palmQuasiadiabaticPathIntegral2018}, which assumes commuting coupling operators ($[S^1,S^2]=0$), induces a first order error in the time step. Note that, due to limiting floating-point precision and the dependence of the depth of the uniTEMPO network on the time-step size, the time step cannot be decreased to arbitrarily small values. The non-trivial Trotter corrections in Eq.~\eqref{eq:infl_gauss0} are thus crucial to achieve convergence.

\begin{figure}
    \centering
    \includegraphics[width=105.5pt]{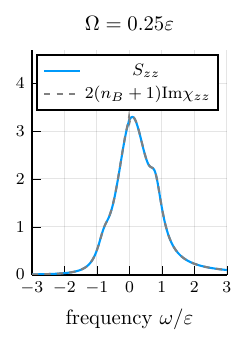}
    \includegraphics[width=105.5pt]{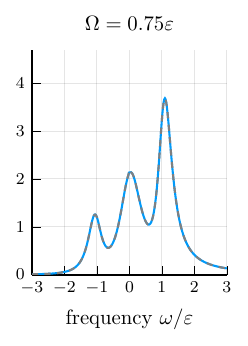}\vspace{-10pt}
    \caption{Symmetrized power spectral density \eqref{eq:PSDzz} for the spin boson model with Jaynes-Cummings-type coupling \eqref{eq:SB_JC} with $H_\mathrm{sys}=\varepsilon\sigma_z/2+\Omega\sigma_x/2$ and an ohmic bath \eqref{eq:bcff_ohmic} $\omega_c=5\varepsilon,\,\beta\varepsilon=1,\alpha=0.05$. The two plots show spectra for different values of $\Omega$ and their fitting to the fluctuation-dissipation relation \eqref{eq:KMS}. Simulations were performed at $\delta t\,\varepsilon=0.05$ and bond dimension $\chi=170$ ($N_c=65536$).}
    \label{fig:KMS}
\end{figure}

As a second example we consider a spin-boson model with Jaynes-Cummings type coupling to a proper thermal reservoir
\begin{equation}\label{eq:SB_JC}
    H=H_\mathrm{sys} + \sum_k h_k (\sigma_+ b_k+\sigma_- b_k^\dagger) +\sum_k \omega_k b_k^\dagger b_k.
\end{equation}
One has again $S^1=\sigma_x$ and $S^2=\sigma_y$ and the functions $\alpha_\pm$ are given as follows
\begin{equation}\label{eq:JC_bcf2}
    \alpha_\pm(t)=\int\limits_0^\infty \diff\omega J(\omega)\left((1+n_\mathrm{B}(\omega))\eul^{-\ii\omega t}\pm n_\mathrm{B}(\omega)\eul^{\ii\omega t}\right).
\end{equation}
For the spectral density $J(\omega)$ we consider an Ohmic bath with exponential high-frequency cutoff
\begin{equation}\label{eq:bcff_ohmic}
J(\omega) = \alpha \omega \eul^{-\omega/\omega_c}.
\end{equation}
We compute dynamical response functions in order to verify thermalization of the system. We consider here specifically the following symmetrized power spectral density
\begin{equation}\label{eq:PSDzz}
S_{zz}(\omega) = 2 \mathrm{Re} \int_0^\infty \diff t\braket{\sigma_z(t)\sigma_z(0)}\eul^{\ii\omega t}
\end{equation}
and the spin susceptibility
\begin{equation}\label{eq:suscep}
    \chi_{zz}(\omega)=\ii\int_0^\infty \diff t\braket{[\sigma_z(t),\sigma_z(0)]}\eul^{\ii\omega t},
\end{equation}
where $\braket{...}$ denotes the expectation value with respect to the stationary state. These linear response functions can be extracted very efficiently using a spectral decomposition obtained from diagonalizing the uniTEMPO propagator \cite{garbelliniUniformProcessTensor2026}. In equilibrium settings, such as defined by the Ohmic bath at a given inverse temperature $\beta$, the fluctuation-dissipation relation connects the power spectral density and the susceptibility \cite{kuboFluctuationdissipationTheorem1966}
\begin{equation}\label{eq:KMS}
S_{zz}(\omega)=2\big(1+n_\mathrm{B}(\omega)\big) \mathrm{Im}\chi_{zz}(\omega),
\end{equation}
indicating that the steady state of the system is consistent with the thermal state of the full Hamiltonian. This relation is an excellent test because our method does not structurally enforce equilibrium dynamics, and can also be used in non-equilibrium settings. Thus, the fluctuation dissipation-relation is satisfied only when the dynamics is numerically converged. As seen in Fig.~\ref{fig:KMS}, we find that the left- and right-hand side of \eqref{eq:KMS} indeed match at all frequencies, demonstrating convergence of the stationary response functions as well as thermalization of the system at the bath temperature. 

\section{Emitters coupled to a bosonic lattice}\label{sec:exampl}

\begin{figure}[t]
    \centering
    \includegraphics[trim={0 10pt 0 0}, width=210pt]{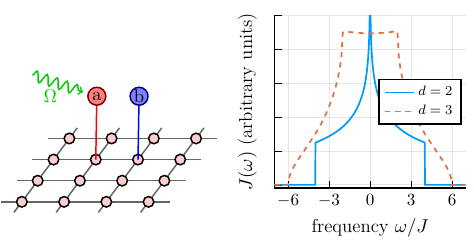}
    \caption{Left: Sketch of the model. Two emitters are coupled to a non-interacting bosonic lattice at different lattice sites. One of the emitters is driven. Right: Local spectral density \eqref{eq:sds} for a cubic lattice in two and three dimensions.}
    \label{fig:sds}
\end{figure}

As a challenging example problem we consider two emitters (two-level systems) $a$ and $b$ coupled locally to a bosonic lattice at different sites $\vec{x}^{a/b}$
\cite{nuominEfficientSimulationOpen2024,papaefstathiouEfficientTensornetworkSimulation2025}, as depicted in Fig.~\ref{fig:sds}. Introducing the local bosonic bath modes $[b_{\vec{x}},b_{\vec{y}}^\dagger]=\delta_{\vec{x},\vec{y}}$ the interaction Hamiltonian reads 
\begin{equation}\label{eq:emitter_model1}
    H_\mathrm{int}= g\sum_{l=a,b} (\sigma^l_- b_{\vec{x}^l}^\dagger + \sigma^l_+b_{\vec{x}^l}),
\end{equation}
and the environment Hamiltonian is
\begin{equation}\label{eq:emitter_model2}
H_\mathrm{env}=-J\sum_{\langle \vec{x},\vec{y}\rangle} (b_{\vec{x}}^\dagger b_{\vec{y}} + b_{\vec{y}}^\dagger b_{\vec{x}}),
\end{equation}
where $\langle \vec{x},\vec{y}\rangle$ denotes a sum over nearest neighbors. For a $d$-dimensional cubic lattice and zero temperature the bath correlation function for any spatial dimension can be expressed explicitly in terms of Bessel functions. To understand better the features of this structured reservoir, it is instructive to consider the local bath spectral density for a single emitter, given as
\begin{equation}\label{eq:sds}
    J(\omega)=\frac{g^2}{(2\pi)^d}\int_\mathrm{BZ} \diff^dk\delta(\omega-\omega(\vec{k})),
\end{equation}
with the lattice dispersion $\omega(\vec{k})=-2J\sum_{i=1}^d \cos(k_i)$. For $d=2$ and $d=3$ the spectral densities are displayed in Fig.~\ref{fig:sds}. The two-dimensional case is generally challenging when the emitter detuning is small, due to the sharp peak in the bath spectrum at $\omega=0$ \cite{papaefstathiouEfficientTensornetworkSimulation2025}, while the spectrum is broad in three dimensions. Moreover, the effective interaction between the emitters mediated through the lattice becomes stronger when the spacial dimension is lower, leading to larger off-diagonal contributions in the bath correlation function.

In the emitter Hamiltonian we assume no direct coupling 
\begin{equation}\label{eq:emitter_model3}
H_\mathrm{sys}=\frac{\Delta}{2}(\sigma_z^{a}+\sigma_z^{b})+\frac{\Omega}{2} \sigma_x^{a}.
\end{equation}
The parameters are the emitter detuning $\Delta$ and a driving strength $\Omega$ for emitter $a$. We place the emitters at neighboring lattice sites $\vec{x}^a-\vec{x}^b=\vec{e}_x$. The emitters are coupled indirectly through the interaction with the lattice.

Besides the highly non-Markovian bath a serious computational hurdle in uniTEMPO for this system is the large physical dimension of the influence functional. For two emitters with two coupling operators each the influence functional has the dimension $4^4=256$ for a physical index ($\mu_n$ in Eq.~\eqref{eq:mpo_if}). For such large physical dimensions the computation can be significantly accelerated by using a low-rank singular value decomposition in the iTEBD contraction. To compute this partial SVD one computes the dominant column space using randomized matrix sketching \cite{hoJuliaMatricesLowRankApproxjlV0522022} in order to reduce the SVD cost by restricting it to the relevant low-rank sector. In addition, we also used the advanced iTEBD contraction scheme proposed in Ref.~\cite{cochinEfficientConstructionTimeinvariant2026} which lowers the computation time in the case of a large system Hilbert space. However, for this particular model we observed that most of the speedup was already realized by the low-rank SVD and filtering of exact degeneracies in the tensors \eqref{eq:b_tens}. With computation times below one hour we can reach bond dimensions of around $\chi\approx 200$ on consumer hardware. Once the semi-group influence functional MPO is constructed, computing dynamics for a given system Hamiltonian requires only matrix-vector multiplications, which is computationally inexpensive at these bond dimensions.

\begin{figure}[t]
    \centering
    \includegraphics[width=225pt]{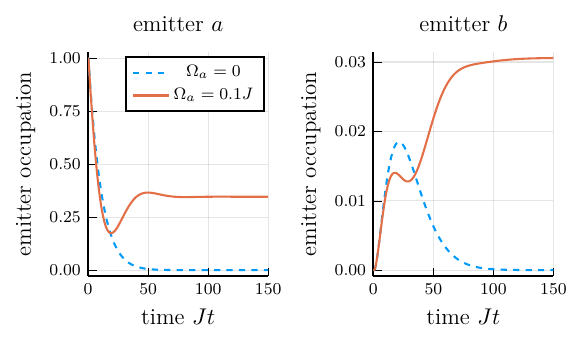}\vspace{-10pt}
    \caption{Dynamics of the emitter occupation ($\sigma_+\sigma_-$) for two emitters coupled to a three-dimensional cubic bosonic lattice at neighboring sites  $g^2=0.1 J^2$, $\Delta=0$ (Eq.~\eqref{eq:emitter_model1},\eqref{eq:emitter_model2},\eqref{eq:emitter_model3}). The emitter $a$ is initially occupied while emitter $b$ is not. Simulation results for $\delta tJ=0.025$ and $\chi=117$ ($N_c=2048$).}
    \label{fig:3d}
\end{figure}

Simulation results for the simpler case $d=3$ are displayed in Fig.~\ref{fig:3d}. We considered both driven and undriven dynamics, where initially the emitter $a$ is occupied while $b$ is empty. Without driving the emitter occupation decays as the system reaches its trivial ground state for long evolution times. The excitation of emitter $a$ partly decays into emitter $b$ leading to a small occupation that likewise decays into the lattice. When emitter $a$ is driven the dynamics is no longer constrained to the single-excitation sector. The emitter population then stabilizes to a finite value. The driving of emitter $a$ also induces a slow buildup of a stationary occupation in emitter $b$.

\begin{figure}[t]
    \centering
    \includegraphics[width=225pt]{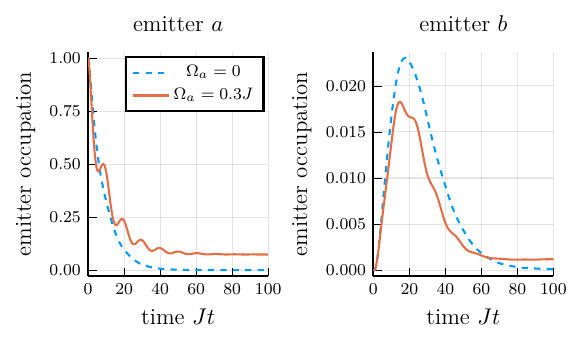}\vspace{-10pt}
    \caption{Dynamics of the emitter occupation ($\sigma_+\sigma_-$) for two emitters coupled to a two-dimensional cubic bosonic lattice at neighboring sites  $g^2=0.1 J^2$, with nonzero detuning $\Delta=0.5J$ (Eq.~\eqref{eq:emitter_model1},\eqref{eq:emitter_model2},\eqref{eq:emitter_model3}). The emitter $a$ is initially occupied while emitter $b$ is not. Simulation results for $\delta tJ=0.05$ and $\chi=357$ ($N_c=131072$).}
    \label{fig:2d}
\end{figure}

For a two-dimensional lattice $d=2$ we were not able to achieve convergence in the case of zero detuning $\Delta=0$ using bond dimensions accessible on a commodity workstation. In appendix \ref{app:converg} we show that sufficient convergence for this resonant case already requires bond dimensions above $\chi=400$ for a single emitter, implying much larger bond dimensions for two emitters. We instead present simulation results for finite detuning in Fig.~\ref{fig:2d}. Finite detuning makes convergence easier because it moves the relevant frequencies involved in the dynamics away from from the low frequency singularity of the local spectral density (depicted in Fig.~\ref{fig:sds}) \cite{windtFermionicMatterwaveQuantum2024,papaefstathiouEfficientTensornetworkSimulation2025}. However, the detuning also suppresses a buildup of occupations in the second emitter because transitions are no longer resonant. This leads to a suppression of the stationary occupation of emitter $b$.

\section{Conclusions}\label{sec:concl}
In this paper we derived a time-discrete influence functional for general Gaussian bosonic baths that is consistent with a second order Trotter decompositon and can therefore be used for long-time simulations of non-Markovian open quantum systems with non-additive environments. In contrast to Ref.~\cite{richter2022Enhanced} the result structurally resembles the original Feynman-Vernon expression and can be directly implemented in the standard QUAPI \cite{makriTensorPropagatorIterative1995} and TEMPO \cite{strathearnEfficientNonMarkovianQuantum2018, jorgensenExploitingCausalTensor2019} algorithms. Here we use specifically the efficient uniform TEMPO algorithm \cite{linkOpenQuantumSystem2024a} to factorize the influence functional as an infinite matrix product operator. We considered an example problem where two emitters are coupled to a non-interacting bosonic lattice at zero temperature. This shows a central benefit of the tensor network influence functional approach. The efficient compression of the environment achieved through a tensor network contraction makes it possible to implement highly structured reservoirs with multiple coupled subsystems, a setup that is not straightforward to realize via other approaches such as HEOM \cite{tanimuraNumericallyExactApproach2020} or pseudomodes \cite{mascherpaOptimizedAuxiliaryOscillators2020, huangCoupledLindbladPseudomode2026}, and that is demanding for chain mappings \cite{papaefstathiouEfficientTensornetworkSimulation2025}. Currently the main bottleneck of the method is the unfavorable numerical scaling with the system Hilbert-space size \cite{cygorekTreelikeProcessTensor2024} combined with very long memory times, as encountered in 2D reservoirs. Large bond dimensions are required to capture long-time memory effects accurately whereas large system sizes restrict the achievable bond dimensions. Note, however, that the influence functional depends only on the space directly coupled to the bath and can be included in a many-body simulation with local independent baths \cite{fuxTensorNetworkSimulation2023}. Overall, the uniTEMPO algorithm performs very well even for highly coherent baths and it is numerically inexpensive for small systems. The generalization developed in this paper allows the method to be applied to problems with arbitrary linear coupling, including the rotating wave coupling ubiquitous in quantum optics. 

\begin{acknowledgments} 
The author acknowledges supported by the Deutsche Forschungs\-gemein\-schaft (DFG, German Research Foundation) via the Research Unit FOR 5688 (Project No. 521530974). \end{acknowledgments}

\bibliography{bib}

\newpage
\appendix
\begin{widetext}

\section{Derivation of the influence functional}\label{app:deriv}

We evaluate the expression \eqref{eq:full_if} analytically in order to derive \eqref{eq:infl_gauss} and \eqref{eq:infl_gauss0}. As a first elementary step we compute the action of a single unitary from $\eqref{eq:full_if}$ on a coherent bath state $\ket{z}=\exp(\sum_\lambda b_\lambda^\dagger z_\lambda)\ket{0}$. Since the unitaries are simple displacement operators, they map coherent states to coherent states
\begin{equation}
    U ^{l}_i(n\delta t,(n-1)\delta t)\ket{z_0}=M\ket{z}.
\end{equation}
The left-hand side can be determined analytically. Defining
\begin{equation}
    U ^{l}_i(t,(n-1)\delta t)\ket{z_0}=M(t)\ket{z(t)}
\end{equation}
we obtain the evolution equation
\begin{equation}
    {\dot M(t)}\ket{z(t)}+\sum_\lambda M(t) \dot z_\lambda(t)\partial_{z_\lambda} \ket{z(t)}=-\ii S_i^l\sum_\lambda M(t)(g^l_\lambda z_\lambda(t)\eul^{-\ii\omega_\lambda t}+g^{l*}_\lambda \eul^{\ii\omega_\lambda t}\partial_{z_\lambda}) \ket{z(t)}.
\end{equation}
Comparing both sides yields differential equations for $z$ and $M$ 
\begin{equation}
    \begin{split}
         \dot{z}_\lambda(t) = -\ii S_i^lg^{l*}_\lambda \eul^{\ii\omega_\lambda t},\,\qquad \frac{\diff}{\diff t}\ln M(t)=-\ii S_i^l\sum_\lambda (g^l_\lambda z_\lambda(t)\eul^{-\ii\omega_\lambda t})
    \end{split},
\end{equation}
with formal solution
\begin{equation}
z_\lambda(t)=\int_{(n-1)\delta t}^t\diff s(-\ii S_i^lg^{l*}_\lambda \eul^{\ii\omega_\lambda s})+z_{0\lambda} ,
        \qquad \ln M(t)=-\int_{(n-1)\delta t}^{t}\diff t \ii S_i^l\sum_\lambda g^l_\lambda z_\lambda(t)\eul^{-\ii\omega_\lambda t}.
\end{equation}
Inserting the first equation in the second gives
\begin{equation}
    \ln M=-\ii \int_{(n-1)\delta t}^{n\delta t}\diff t S_i^l\sum_\lambda g^l_\lambda z_{0\lambda}\eul^{-\ii\omega_\lambda t}
    -\int_{(n-1)\delta t}^{n\delta t}\diff t \int_{(n-1)\delta t}^{t}\diff sS_i^l S_i^l\sum_\lambda g^l_\lambda g^{l*}_\lambda \eul^{-\ii\omega_\lambda (t-s)},
\end{equation}
\begin{equation}
    z_\lambda=\int_{(n-1)\delta t}^{n\delta t}\diff s(-\ii S_i^lg^{l*}_\lambda \eul^{\ii\omega_\lambda s})+z_{0\lambda}.
\end{equation}
This step must be repeated $L$ times for a single time step $n$ in \eqref{eq:full_if}. We assume an odd time step where the Trotter order is
\begin{equation}
    U^{L}_{i_n^L}(n\delta t,(n-1)\delta t)\cdots U^{1}_{i_n^1}(n\delta t,(n-1)\delta t)\ket{z_0}=M\ket{z},
\end{equation}
giving 
\begin{equation}
    z_\lambda = \sum_{l=1}^L \int_{(n-1)\delta t}^{n\delta t}\diff s(-\ii S_{i_n^l}^lg^{l*}_\lambda \eul^{\ii\omega_\lambda s})+z_{0\lambda}.
\end{equation}
Note that for $z_\lambda$ the order of the unitaries is irrelevant because the displacements commute. However, the order is relevant in the prefactor $M$
\begin{equation}
\begin{split}
    \ln M=&-\ii \sum_{l=1}^L\int_{(n-1)\delta t}^{n\delta t}\diff t S_i^l\sum_\lambda g^l_\lambda z_{0\lambda}\eul^{-\ii\omega_\lambda t}\\& 
    -\sum_{l=1}^L\int_{(n-1)\delta t}^{n\delta t}\diff t \int_{(n-1)\delta t}^{t}\diff sS_{i_n^l}^l S_{i_n^l}^l\alpha^{ll}(t-s)\\& 
    -\sum_{l=1}^L\sum_{o=1}^{L}\int_{(n-1)\delta t}^{n\delta t}\diff t \int_{(n-1)\delta t}^{n\delta t}\diff s \Theta_{l-o} S_{i_n^l}^l S_{i_n^o}^{o}\alpha^{lo}(t-s),
\end{split}
\end{equation}
where $\Theta_n=0$ if $n\leq 0$ and $\Theta_n=1$ else.
For an even time step the order is swapped which would give $\Theta_{o-l}$ instead of $\Theta_{l-o}$ in the last line. We thus define the following matrix for a convenient notation
\begin{equation}
    R^n_k=\begin{cases}
        \Theta_k &n \text{ even}\\
        \Theta_{-k} &n \text{ odd}
    \end{cases}.
\end{equation}
To obtain the full result we iterate the previous steps $N$ times with $z_0=0$ as initial state. This yields
\begin{equation}
    z_\lambda = \sum_{n=1}^N\sum_{l=1}^L \int_{(n-1)\delta t}^{n\delta t}\diff s(-\ii S_{i_n^l}^lg^{l*}_\lambda \eul^{\ii\omega_\lambda s}),
\end{equation}
\begin{equation}
\begin{split}
    \ln M=&-\sum_{n=1}^{N}\sum_{m=1}^{n-1}\sum_{l=1}^L\sum_{o=1}^L \int_{(n-1)\delta t}^{n\delta t}\diff t \int_{(m-1)\delta t}^{m\delta t}\diff s S_i^l S_{i_m^o}^o \alpha^{lo}(t-s)\\& 
    -\sum_{n=1}^N\sum_{l=1}^L\int_{(n-1)\delta t}^{n\delta t}\diff t \int_{(n-1)\delta t}^{t}\diff sS_{i_n^l}^l S_{i_n^l}^l\alpha^{ll}(t-s)\\& 
    -\sum_{n=1}^N\sum_{l=1}^L\sum_{o=1}^{L}R^n_{o-l}\int_{(n-1)\delta t}^{n\delta t}\diff t \int_{(n-1)\delta t}^{n\delta t}\diff s S_{i_n^l}^l S_{i_n^o}^{o}\alpha^{lo}(t-s).
\end{split}
\end{equation}
For the influence functional we also need the overlap of forward and backward evolved coherent states which generates the forward and backward path mixing terms. Specifically, the influence functional can be obtained via Eq.~\eqref{eq:full_if} as
\begin{equation}
\mathcal{F}= \bra{{0}}\cdots U^{1\,\dagger}_{j_{N}^L}(N\delta t,N\delta t- \delta t)U^{1}_{i_N^L}(N\delta t,N\delta t-\delta t)\cdots\ket{{0}}=\bra{z'}(M')^*M\ket{z}=\exp(\ln M +\ln {M'}^*+z{z'}^*),
\end{equation}
where the priming $z,M\rightarrow z',M'$ amounts to replacing the forward $i_n^l$ indices with the backwards $j_n^l$ indices. The coherent state overlap term gives
\begin{equation}
    z {z'}^*=\sum_\lambda z_\lambda {z_\lambda'}^*=\sum_{n=1}^N\sum_{m=1}^N\sum_{l=1}^L\sum_{o=1}^L \int_{(n-1)\delta t}^{n\delta t}\diff t\int_{(m-1)\delta t}^{m\delta t}\diff s S_{i_n^l}^lS_{{j_m^o}}^o \alpha^{ol}(s-t).
\end{equation}
This expression can be split into three parts
\begin{equation}
\begin{split}
        z {z'}^*=&+\sum_{n=1}^N\sum_{m=1}^{n-1}\sum_{l=1}^L\sum_{o=1}^L \int_{(n-1)\delta t}^{n\delta t}\diff t\int_{(m-1)\delta t}^{m\delta t}\diff s S_{i_n^l}^lS_{{j_m^o}}^o \alpha^{ol}(s-t)\\
        &+\sum_{n=1}^N\sum_{m=n+1}^N\sum_{l=1}^L\sum_{o=1}^L \int_{(n-1)\delta t}^{n\delta t}\diff t\int_{(m-1)\delta t}^{m\delta t}\diff s S_{i_n^l}^lS_{{j_m^o}}^o \alpha^{ol}(s-t)\\
        &+\sum_{n=1}^N\sum_{l=1}^L\sum_{o=1}^L \int_{(n-1)\delta t}^{n\delta t}\diff t\int_{(n-1)\delta t}^{n\delta t}\diff s S_{i_n^l}^lS_{{j_n^o}}^o \alpha^{ol}(s-t).
\end{split}
\end{equation}
Moreover, we can use $\alpha^{ol}(-t)=(\alpha^{lo}(t))^*$ 
to further rewrite the boundary term (last line) as
\begin{equation}
    \int_{(n-1)\delta t}^{n\delta t}\diff t \int_{(n-1)\delta t}^{n\delta t}\diff s\alpha^{lo}(t-s)
    =\int_{0}^{\delta t}\diff t \int_{0}^{t}\diff s \alpha^{lo}(t-s)+\int_{0}^{\delta t}\diff t \int_{0}^{t}\diff s \alpha^{ol*}(t-s).
\end{equation}
Carefully collecting all terms in the exponent of the influence functional $(\ln M + \ln {M'}^{*} + z {z'}^*)$ we obtain the end result \eqref{eq:infl_gauss}, \eqref{eq:infl_gauss0}.

\section{Convergence of the 2D bath problem}\label{app:converg}

Here we show that the convergence of the emitter example for the case of a two-dimensional lattice and for zero detuning $\Delta=0$ requires large bond dimensions. We consider specifically the case of a single emitter, for which larger bond dimensions are achievable with low numerical effort. If the driving is zero $\Omega_a=0$ then the system has at most one excitation such that the exact solution can be easily obtained numerically. 

Convergence of the single particle dynamics in the case of zero and nonzero detuning is displayed in Fig.~\ref{fig:2dsingle}. For zero detuning and small bond dimensions the uniTEMPO simulation results in a non-positive density matrix for the emitter, visible through negative emitter occupations. We find systematic convergence towards the exact solution as the bond dimension is increased. However, relatively large bond dimensions are necessary that would be computationally very demanding in a two-emitter simulation. For finite detuning the convergence is generally faster, and simulations with lower bond dimension already achieve high accuracy. 

Note that for a finite driving $\Omega\neq 0$ the dynamics is no longer constrained to the single excitation sector and an exact reference solution can no longer be obtained from the single particle dynamics. 

\begin{figure}[h]
    \centering
    \includegraphics[width=150pt]{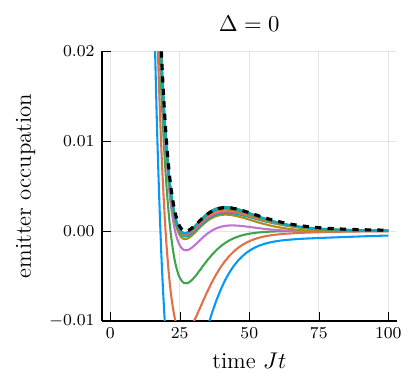}
    \includegraphics[width=150pt]{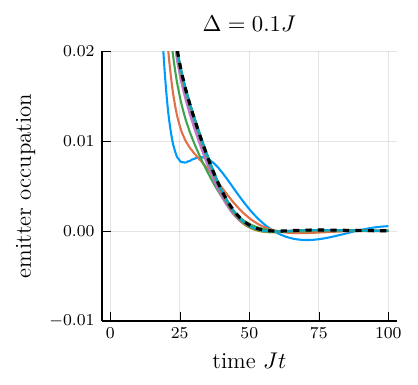}
    \includegraphics[width=150pt]{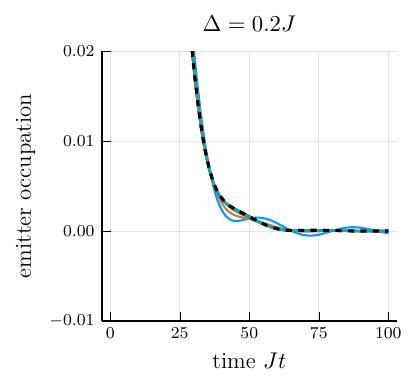}\vspace{-10pt}

    \caption{Dynamics of the emitter occupation ($\sigma_+\sigma_-$) for an initially occupied single emitter coupled to a two-dimensional cubic bosonic lattice $g^2=0.1 J^2$ (as in Eq.~\eqref{eq:emitter_model1},\eqref{eq:emitter_model2},\eqref{eq:emitter_model3}, but with emitter $a$ only, $\Omega=0$) for zero detuning (left) and finite detuning (center, right). The dashed line is the exact single-particle solution. Simulation results for $\delta tJ=0.05$ and varying bond dimensions $\chi=52, 83, 141, 188, 251, 279, 301, 322, 349, 370, 408$. Simulation results with larger bond dimensions are closer to the exact solution. For finite detuning the convergence with respect to the bond dimension is faster.}
    \label{fig:2dsingle}
\end{figure}

\end{widetext}

\end{document}